\documentclass[12pt]{article}
\usepackage{epsfig}
\usepackage{amsmath,amssymb,amsfonts,latexsym,cancel,slashed}
\usepackage[mathscr]{euscript}
\begin{document} 
\begin{titlepage}

\renewcommand{\thefootnote}{\fnsymbol{footnote}}
\vspace*{2cm}

\begin{center}
{\Large \bf Polarized semi-inclusive electroweak structure \\ functions at next-to-leading-order}
\end{center}

\par \vspace{2mm}
\begin{center}

{\bf Daniel de Florian and Yamila Rotstein Habarnau}

Departamento de F\'isica, FCEyN, Universidad de Buenos Aires,\\
(1428) Pabell\'on 1 Ciudad Universitaria, Capital Federal, Argentina

\vspace{5mm}

\end{center}

\par \vspace{2mm}
\begin{center} {\large \bf Abstract} \end{center}
\begin{quote}
\pretolerance 10000

We present a next-to-leading order (NLO) computation of the full set of polarized and unpolarized electroweak semi-inclusive DIS (SIDIS) structure functions, 
whose knowledge is crucial for a precise extraction of polarized parton distributions. 
We focus on the phenomenology of the polarized structure functions for the kinematical conditions that could be reached in an Electron-Ion-Collider.

We show that the NLO corrections are sizeable, particularly in the small-$x$ range.
We test the sensitivity of these structure functions on certain quark distributions
and compare it to the situation of inclusive DIS and electromagnetic SIDIS. 

\end{quote}

\vspace*{\fill}
\begin{flushleft}
October 2012

\end{flushleft}
\end{titlepage}

\setcounter{footnote}{1}
\renewcommand{\thefootnote}{\fnsymbol{footnote}}

\section{Introduction}

Understanding how the nucleon spin is composed of the angular momenta and spins of its constituents (quarks and gluons) has been 
a defining question in hadron structure for a long time. Since it was found that little of the proton spin is carried by the quarks and anti-quarks spins, 
several experiments have measured with increasing precision observables which are sensitive to quark and gluon polarizations in the nucleon. 
Such experimental progress was matched by advancements in theoretical precision and phenomenological analyses of data \cite{review}.

The spin structure of a nucleon can be described by the (anti)quark and gluon polarized parton distribution functions (pPDFs), defined by 
\begin{equation}
\Delta f_j(x,Q^2)  \equiv f_j^+(x,Q^2) \; - \; f_j^-(x,Q^2),
\label{qdef}
\end{equation}
where $f_j^+(x,Q^2)$ ($f_j^-(x,Q^2)$) denotes the distribution of a parton $j$ with positive (negative) helicity in 
a nucleon with positive helicity, as a function of momentum fraction $x$ and scale $Q$.
Its first momentum, that is to say the integral 
$\Delta f_j^1(Q^2)\equiv\int_0^1 \Delta f_j(x,Q^2) dx$,  
directly measures the spin contribution of the parton $j$ to the proton spin.

The most complete global fit includes all available data taken in spin-dependent deep inelastic scattering
(DIS), semi-inclusive DIS (SIDIS) with identified pions and kaons, and 
proton-proton collisions. They allow  us to extract sets of pPDFs consistently at next-to-leading order (NLO) in the strong coupling constant 
along with estimates of their uncertainties \cite{DSSV1,DSSV2}.

Unlike unpolarized PDF fits, where a clear separation of different quark flavours is possible
by the use of inclusive charged-current DIS data, differences in polarized quark and anti-quark 
distributions are determined exclusively from SIDIS data and hence require knowledge of the hadronization mechanism, encoded in 
non-perturbative fragmentation functions (FFs). Pion FFs are rather well known, but uncertainties for kaon FFs are much larger. 
Significant progress on the quality of fits of FFs
is expected once data from $B$ factories and the LHC become available.

Despite the impressive experimental and theoretical progress made in the last years, many fundamental questions
related to the proton spin structure still remain unanswered.
One of the main problems is that present fixed-target experiments suffer from their very limited kinematic coverage 
in $x$ and $Q^2$. The kinematic range could be extended in an Electron-Ion Collider (EIC), whose set-up is being currently considered \cite{EIC}, 
and thus SIDIS measurements  would allow us to extract 
$\Delta u$, $\Delta \bar{u}$, $\Delta d$, $\Delta \bar{d}$, $\Delta s$, and $\Delta \bar{s}$ with much higher precision.
Furthermore, this new collider would present the opportunity to perform DIS and SIDIS measurements via charged and neutral electroweak currents, 
which permits to access polarized electroweak structure functions \cite{EWSF} that depend on various combinations of polarized quark PDFs and provide
an effective way of disentangling different quark flavours.

Experimentally, one has access to the asymmetries, which depend on both the polarized an unpolarized structure functions ($g_i$ and $F_i$, respectively). 
The unpolarized SIDIS structure functions are well known at next-to-leading order (NLO) in perturbative Quantum-Chromodynamics (QCD) \cite{AltarelliSIDIS,deFlorianLambda}, 
as well as the  electromagnetic polarized ones \cite{deFlorianLambda}.
However, polarized SIDIS structure functions for charged current and $Z$ boson exchange are only known to leading-order accuracy (LO) \cite{Anselmino}.
In general, a LO calculation only captures the main features but does not provide a quantitative description of the process. 
It is then necessary to know the NLO QCD corrections to both the unpolarized and polarized structure functions in order to 
extract reliable information on the parton distribution functions.

In this work we present a NLO computation of both polarized and unpolarized SIDIS
structure functions.
We focus on the electroweak ones, which are particularly useful to achieve a full flavour separation.
The paper is organized as follows. In section \ref{sec:LO} we establish the definition of electroweak (un)polarized structure functions and present their LO expressions. 
In section \ref{sec:NLO} we explain the main features of the computation of the structure functions at NLO, focusing on the new results for the polarized case. 
In section \ref{sec:phenomenology} we present some phenomenological results and analyse the relevance of the NLO description of polarized SIDIS.
Finally, we present our conclusions in section \ref{sec:conclusions}.


\section{SIDIS electroweak structure functions \label{sec:LO}}

In this section we establish the definition of the electroweak structure functions at lowest order in perturbation theory.
We focus, in this and the following sections, on the semi-inclusive case, 
leaving some comments on the totally inclusive DIS to Appendix \ref{sec:ap_DIS}. For more details on DIS notation, we refer to \cite{PDG}.

In lowest-order perturbation theory of electroweak interactions, the cross section for the scattering of polarized
leptons on polarized nucleons and the consequent observation of a hadron $H$ in the final state, 
can be expressed in terms of the product of a leptonic and a hadronic tensor as 
\begin{equation}
\frac{d\sigma^H}{dx\,dy\,dz}=\frac{2\pi\, y \, \alpha^2}{Q^4} \sum_{j} \eta_j L_j^{\mu\nu}W^{H,j}_{\mu\nu},
\end{equation}
where $x$ and $y$ denote the usual DIS variables, $-q^2=Q^2=S x y$, \mbox{$x=Q^2/(2P\cdot q)$} ($q$ being the electroweak current four-momentum, 
and $S$ the center-of-mass energy squared of the lepton-nucleon system) 
and $z=P_H\cdot P/P\cdot q$ the scaling variable representing the momentum fraction taken by the hadron $H$
\footnote{We point out that we concentrate here in the 
current fragmentation region, and thus cuts over the $z$ variable must be imposed (typically, $z>0.1$). For a description of this process
in the target fragmentation region, see Ref. \cite{fracture}.}.
For neutral-current (NC) processes, the sum runs over $j = \gamma$, $Z$ and $\gamma Z$,  representing
photon and $Z$ exchange and the interference between them respectively, while for charged-current (CC)
processes the interaction occurs only via the exchange of a  $W$  boson ($j = W$). 
The tensor $L^{\mu\nu}$ is associated with the coupling of the exchanged boson to the leptons, 
and the hadronic tensor $W_{\mu\nu}^H$ describes the interaction of the appropriate electroweak
currents with the target nucleon and the subsequent hadronization of partons into $H$.
The factors $\eta_j$ denote the ratios of the corresponding propagators and
couplings to the photon propagator and coupling as
\begin{equation}
\eta_\gamma = 1;\quad \eta_{\gamma Z} = \left( \frac{G_F M_Z^2}{2\sqrt{2} 
\pi\alpha} \right) \left( \frac{Q^2}{Q^2 + M_Z^2} \right);\quad \eta_Z = 
\eta^2_{\gamma Z}; \quad \eta_W = \frac{1}{2} 
\left( \frac{G_F M_W}{4\pi\alpha} \frac{Q^2}{Q^2 + M_W^2}\right)^2 .
\label{eq:eta}
\end{equation}

The unpolarized and polarized SIDIS structure functions ($F_i^H$ and $g_i^H$ respectively) are defined in terms of the hadronic tensor \cite{PDG}
\begin{align}
W^{H,j}_{\mu\nu}=&\left( -g_{\mu\nu}+\frac{q_{\mu}q_{\nu}}{q^2}\right) \; F_1^{H,j}(x,z,Q^2) 
		 + \frac{\hat{P}_\mu \, \hat{P}_\nu}{P\cdot q} \; F_2^{H,j}(x,z,Q^2) 
		 -i\epsilon_{\mu\nu\alpha\beta} \frac{q^{\alpha}P^{\beta}}{2P\cdot q} \; F_3^{H,j}(x,z,Q^2) \nonumber \\
		 &+i\epsilon_{\mu\nu\alpha\beta}\frac{q^{\alpha}}{P\cdot q} \left[ S^{\beta} \; g_1^{H,j}(x,z,Q^2) 
		 + \left(S^{\beta}-\frac{S\cdot q}{P\cdot q}\;P^{\beta}\right) \; g_2^{H,j}(x,z,Q^2) \right] \nonumber \\
		 &+\frac{1}{P\cdot q} \left[ \frac{1}{2} \left( \hat{P}_\mu \; \hat{S}_\nu + \hat{S}_\mu \; \hat{P}_\nu \right ) 
		 - \frac{S\cdot q}{P\cdot q} \; \hat{P}_\mu \; \hat{P}_\nu \right] g_3^{H,j}(x,z,Q^2) \nonumber \\
		 &+ \frac{S\cdot q}{P\cdot q} \left[ \frac{\hat{P}_\mu \, \hat{P}_\nu}{P\cdot q} g_4^{H,j}(x,z,Q^2) 
		 + \left(-g_{\mu\nu}+\frac{q_{\mu}q_{\nu}}{q^2}\right) g_5^{H,j}(x,z,Q^2) \right],
\label{eq:Wmunu}
\end{align}
where $P$ and $S$ denote the nucleon momentum and spin four-vectors respectively and  $\hat{P}$, $\hat{S}$ are
\begin{align}
\hat{P}_{\mu}= P_{\mu}-\frac{P\cdot q}{q^2}\;q_{\mu} \ , \  \hat{S}_{\mu}= S_{\mu}-\frac{S\cdot q}{q^2}\;q_{\mu} \; .
\end{align}

The spin-averaged SIDIS cross section for $e^\pm N$ scattering and the subsequent production of a
hadron $H$ in the current fragmentation region, for $Q^2\gg M^2$ (where $M$ is the mass of the 
nucleon), is given in terms of the unpolarized structure functions by
\begin{equation}
\frac{d\sigma^{H,i}}{dx \,dy \,dz} = \frac{2\pi\alpha^2}{x\,y\,Q^2} \eta^i \Bigg[
\big[ 1+(1-y)^2 \big]\, 2x\, F_1^{H,i} \mp \big[ 1-(1-y)^2 \big] xF_3^{H,i} +(1-y) \, 2 \, F_L^{H,i} \Bigg] \,\, ,
\label{eq:sigma_nopol}
\end{equation}
where $i$ corresponds to NC or CC. The longitudinal structure function is defined as $F_L^H=F_2^H-2\, x\, F_1^H$ and 
vanishes at lowest order according to the Callan-Gross relation \cite{Callan}.  $\eta^{NC}=1$, while
$ \eta^\mathrm{CC} = (1\pm\lambda)^2\eta_W$ ($\eta_W$ defined in Eq. \eqref{eq:eta}), where $\lambda=\pm 1$ represents the electron/positron helicity. 
(For incoming neutrinos, it is $ \eta^\mathrm{CC} = 4\eta_W$  instead.) Here and in what follows, the sign $\pm$ refers to the lepton charge.  

The NC structure functions can be obtained as the sum of the photon, 
$Z$, and interference contributions:
\begin{equation}
F_{1,L}^\mathrm{H,NC} = F_{1,L}^{H,\gamma} - (g_V^e \pm \lambda g_A^e) \, \eta_{\gamma Z} \,
F_{1,L}^{H,\gamma Z} + (g_V^e{}^2 + g^e_A{}^2 \pm 2\,\lambda \,g_V^e \,g_A^e) \,\eta_Z \,F_{1,L}^{H,Z}
\end{equation}
and 
\begin{equation}
x \, F_3^\mathrm{H,NC} = -(g_A^e\pm\lambda \, g_V^e) \, \eta_{\gamma Z} \,x \,  F_3^{H,\gamma Z} + 
[2 \, g_V^e \, g_A^e \pm \lambda \, (g_V^e{}^2 + g_A^e{}^2)] \, \eta_Z \, x \, F_3^{H,Z},
\end{equation}
with $g_V^e = -\frac{1}{2} + 2\sin^2\theta_W$, $g_A^e = -\frac{1}{2}$.

For the case of a polarized target, the difference $\Delta \sigma$ of cross sections for 
the two nucleon helicity states is
\begin{equation}
\frac{d\Delta\sigma^{H,i}}{dx\,  dy\, dz} = \frac{8\pi\alpha^2}{x\,y \, Q^2} \eta^i \Bigg[
\big[ 1+(1-y)^2\big] x\, g_5^{H,i} \pm \big[ 1-(1-y)^2\big] x\, g_1^{H,i} + (1-y)   \,  g_L^{H,i} \Bigg],
\label{eq:sigma_pol}
\end{equation}
where again $i$ corresponds to NC or CC and where $g_L^H = g_4^H - 2xg_5^H$. 
Like $F_L$, the latter vanishes at leading order. 
The NC spin dependent structure functions are 
\begin{eqnarray}
g_1^\mathrm{H,NC} & = & \pm \lambda \, g_1^{H,\gamma} - (g_A^e \pm \lambda g_V^e) \, 
\eta_{\gamma Z} \, g_1^{H,\gamma Z} + ( 2 \, g_V^e \, g_A^e \pm \lambda(g_V^e{}^2 + 
g^e_A{}^2) ) \,  \eta_Z \, g_1^{H,Z}, \nonumber\\
g_{5,L}^\mathrm{H,NC} & = & -(g_V^e \pm \lambda g_A^e) \, \eta_{\gamma Z} \,  
g_{5.L}^{H,\gamma Z} + (g_V^e{}^2 + g^e_A{}^2 \pm 2 \lambda \, g_V^e \, g_A^e) \, \eta_Z  \, 
g_{5,L}^{H,Z}.
\end{eqnarray}

In the quark-parton model, at leading-order, contributions to the SIDIS structure functions $F_i$ and $g_i$ can be expressed in terms 
of the quark (p)PDFs and the hadron FFs. In the NC case, these functions are
\begin{eqnarray}
\left[ F_1^{H,\gamma},\, F_1^{H, \gamma Z},\, F_1^{H,Z} \right] & = & 
\frac{1}{2} \sum_{q} \left[ e_q^2,\, 2 \, e_q \, g_V^q, \, g_V^q{}^2 + g_A^q{}^2 \right] 
(q \, D_q^H+\bar{q}\, D_{\bar{q}}^H) \;, \nonumber\\
\left[ F_3^{H,\gamma},\, F_3^{H,\gamma Z},\, F_3^{H,Z} \right] & = & 
\sum_{q} \left[ 0,\, 2\,e_q \, g_A^q,\, 2\, g_V^q \, g_A^q \right] (q \, D_{q}^H -\bar{q} \, D_{\bar{q}}^H )\;, \nonumber\\
\left[ g_1^{H,\gamma},\, g_1^{H,\gamma Z},\, g_1^{H,Z }\right] & = & 
\frac{1}{2} \sum_{q} \left[ e_q^2,\, 2 \, e_q \, g_V^q,\, 
g_V^q{}^2 + g_A^q{}^2 \right] (\Delta q \, D_{q}^H +\Delta\bar{q}  \, D_{\bar{q}}^H ), \nonumber\\[2mm]
\left[ g_5^{H,\gamma},\, g_5^{H,\gamma Z},\, g_5^{H,Z} \right] & = & 
\sum_{q} \left[ 0, \, e_q \, g_A^q,\, g_V^q \, g_A^q \right] 
(\Delta \bar{q} \, D_{\bar{q}}^H -\Delta q  \, D_{q}^H ) \, \, .
\label{eq:sfun_LO_NC}
\end{eqnarray}
where $e_q$ is the fractional electric charge of the quark, 
$g_V^q = \pm \frac{1}{2}-2 e_q \, \sin^2\theta_W$, and 
$g_A^q = \pm\frac{1}{2}$, with the $+$ sign for up-type quarks and the 
$-$ sign for down-type quarks. 
In the CC case, since the $W$ boson interacts only with certain flavours, we have
(assuming four active flavours):
\begin{eqnarray}
F_1^{H,W^-}  &=&  u \, D_d^H+ \bar{d} \, D_{\bar{u}}^H + \bar{s} \, D_{\bar{c}}^H + c \, D_{s}^H   , \nonumber\\ 
F_3^{H,W^-}  &=&  2(u \, D_d^H - \bar{d} \, D_{\bar{u}}^H - \bar{s} \, D_{\bar{c}}^H + c \, D_{s}^H)  ,\nonumber\\
g_1^{H,W^-}  &=&   \Delta u \, D_d^H + \Delta \bar{d} \, D_{\bar{u}}^H + \Delta \bar{s} \, D_{\bar{c}}^H + \Delta c \, D_{s}^H  , \nonumber\\ 
g_5^{H,W^-}  &=&  -\Delta u \, D_d^H + \Delta \bar{d} \, D_{\bar{u}}^H + \Delta \bar{s} \, D_{\bar{c}}^H - \Delta c \, D_{s}^H \,\, .
\label{eq:sfun_LO_CC}
\end{eqnarray}
For $W^+$ exchange, one should replace $u\leftrightarrow d$ and 
$s\leftrightarrow c$.


\section{Next-to-leading order \label{sec:NLO}}
\label{chap:NLO}

Assuming factorization, the polarized SIDIS structure 
functions $g_{1,5,L}^{H,V}$\footnote{$V$ standing for $\gamma$, $Z$, $\gamma Z$ or $W^{\pm}$.} at a factorization scale $\mu_F$ 
can be expressed as convolutions of non-perturbative pPDFs $\Delta f_k(x,\mu_F^2)$ and FFs $D_j^H(z,\mu_F^2)$ with short-distance 
coefficients $\Delta C_{1,5,L}^{jk,V}(x,z,\mu_F^2)$, which can be evaluated in perturbation theory.
At next-to-leading-order (NLO) in the strong coupling constant $\alpha_s$ these read
\begin{align}
g_i^{H,V}(x,z,Q^2)&= \frac{1}{2} \sum_{q_a,q_b} \Delta \xi_i^{q_a} \left\{\Delta C_{o_i}^V  D_{q_b}^H(z,Q^2) \, \Delta q_a(x,Q^2) 
		    + \frac{\alpha_s(Q^2)}{2\pi} \left[ D_{q_b}^H  \otimes \Delta C_i^{qq,V} \otimes \Delta q_a \right. \right. \nonumber \\  
& \ \ \left. \left. + D_g^H \otimes \Delta C_i^{gq,V} \otimes \Delta q_a  + D_{q_b}^H \otimes \Delta C_i^{qg,V} \otimes \Delta g  \right](x,z,Q^2)\right\}.
\label{eq:sfun_pol}
\end{align}
where  we fix all scales equal to $Q$. 
The sum runs over all contributing partonic channels. That means, for $V=\gamma, \; \gamma Z , \; Z$, 
\begin{equation}
q_a=q_b=u,d,s,c,\bar{u}, \bar{d}, \bar{s} , \bar{c}
\end{equation}
if we consider four active flavours, while for $V=W^-$ only
\begin{equation}
q_a (q_b) =u (d) \; , \; \bar{d} (\bar{u}) \; , \; \bar{s} (\bar{c}) \; , \; {c} ({s})
\label{eq:channels}
\end{equation}
are considered. For $V=W^+$, one should replace $u\leftrightarrow d$ and 
$s\leftrightarrow c$.
The factor  $\Delta \xi_i^{q_a}$ takes the value $\Delta \xi_1^{q_a}=1$ for every $q_a$, 
while it reads $\Delta \xi_5^{q_a}= \Delta \xi_L^{q_a}=1$ if $q_a$ refers to a quark and $\Delta \xi_5^{q_a}= \Delta \xi_L^{q_a}=-1$ for $q_a$ representing an anti-quark. 
In Eq. \eqref{eq:sfun_pol} $\otimes$ denotes the usual convolution:
\begin{align}
(D \otimes C \otimes f)(x,z,Q^2)= \int_x^1\frac{dy}{y}  \int_{z}^1 \frac{d\omega}{\omega} D(\omega) \, C(\frac{x}{y},\frac{z}{\omega},Q^2) \, f(y) \, \,.
\end{align}

According to equations \eqref{eq:sfun_LO_NC} and \eqref{eq:sfun_LO_CC}, which represent the leading-order term of Eq. \eqref{eq:sfun_pol}, 
the coefficients $\Delta C_{o_i}^{V}$ can be written as
\begin{equation}
 \Delta C_{o_1}^{V} = \lambda_V^V \quad ;  \quad \Delta C_{o_5}^{V} = \lambda_A^V \quad ; \quad \Delta C_{o_L}^{V} = 0 \, \, , 
\end{equation}
with
\begin{eqnarray}
\lambda_V^{\gamma}=e_q^2 \; \;, \; \;  \lambda_A^{\gamma}=0   \; \; &,& \; \;  
\lambda_V^{\gamma Z}= 2e_q g_V^q  \; \;, \; \;  \lambda_A^{\gamma Z}=  -2e_q g_A^q \; \;, \\ \nonumber
\lambda_V^{Z}= g_V^q{}^2 + g_A^q{}^2  \; \;, \; \;  \lambda_A^{Z}= - 2g_V^q g_A^q  \; \;&,& \; \;  
\lambda_V^{W}= - \lambda_A^{W}= 2 \;\;.  
\label{eq:lambdaV}
\end{eqnarray}

In order to calculate the NLO coefficients $\Delta C_i^{jk,V}$, we must take into account the one-loop corrections to the partonic process $V+q_a \rightarrow q_b$, 
the real emission $V+q_a \rightarrow q_b + g $ and the box contribution $V+g \rightarrow q_b + \bar{q}_b$. The first two channels contribute to the case with $jk=qq$, 
and the last one to $jk=qg$. The coefficient $\Delta C_i^{gq}$ is obtained from the real emission, considering that the gluon is the hadronizing parton.
At the intermediate stages of the computation divergences appear. In order to regularize them we  use dimensional regularization \cite{regularization,HV},
i.e., we work in a $d$-dimensional space, with $d=4-2\epsilon$. All quarks are considered massless.

Once the matrix element of each channel ${jk}$ has been computed, one must obtain the spin dependent amplitude  
\begin{equation}
 \Delta \left| M^{jk} \right|^2_{\mu\nu}=
\frac{1}{2}\left[ \left| M_+^{jk} \right|^2_{\mu\nu} - \left| M_-^{jk} \right|
^2_{\mu\nu} \right],
\label{eq:DeltaMunu}
\end{equation}
defined in terms of the amplitudes for partons whose polarization is parallel($+$) or anti-parallel($-$) to that of the target. 
The calculation of each term in Eq. \eqref{eq:DeltaMunu} requires projection onto definite helicity states of the incoming particles.  
It is then necessary to make use of the relations
\begin{equation}
 u(p,\lambda)\bar{u}(p,\lambda)=\frac{1+\lambda\, \gamma_5}{2} \slashed{p}
\label{eq:pol_fermion}
\end{equation}
for incoming quarks with helicity $\lambda$ (a similar expression is obtained for anti-quarks) and
\begin{align}
\varepsilon_{\alpha}(p,\lambda_g) \, \varepsilon_{\beta}^*(p,\lambda_g)=\frac{1}{2(1-\epsilon)} \left[ -g_{\alpha\beta}+\frac{p_{\alpha}\, \eta_{\beta}  
+ p_{\beta}\, \eta_{\alpha} }{p.\eta}\right] + \frac{1}{2}  i\, \lambda_g \, \epsilon_{\alpha\beta\rho\sigma} \, \frac{p^{\rho}\, \eta^{\sigma}}{p.\eta} 
\label{eq:pol_gluon}
\end{align}
for incoming gluons with helicity $\lambda_g$, where $\eta$ is an arbitrary light-like momentum, provided that \mbox{ $p.\eta\neq0$}.  
The terms independent of $\lambda$ and $\lambda_g$ in equations \eqref{eq:pol_fermion} and \eqref{eq:pol_gluon} respectively contribute only 
to the unpolarized amplitude (since they cancel out when subtracting the two terms in Eq. \eqref{eq:DeltaMunu}). 
In the last case, the averaging of gluon spins in $d$ dimensions should be performed by dividing by the $d-2=2 (1-\epsilon)$ possible spin orientations, 
as has been made explicit in Eq. \eqref{eq:pol_gluon}.

Once the NLO $\Delta \left| M^{jk} \right|^2_{\mu\nu}$ has been computed, we can obtain each one of the coefficients $\Delta C_i^{jk,V}$ as the finite part of the 
{\it partonic structure function}, defined by 
\begin{equation}
g_i^{jk}=\frac{1}{4 \pi} \int d\Gamma\,\, \Tilde{P}_i^{\mu\nu} \,\, \Delta\left| M^{jk} \right|
^2_{\mu\nu} ,
\label{eq:partonicgi}
\end{equation}
where  $d\Gamma$ is the $d$-dimensional phase-space and the projectors $\tilde{P}_i^{\mu\nu}$ are
\begin{eqnarray}
\tilde{P}_1^{\mu\nu}= -i\, \epsilon^{\mu\nu\rho\sigma} \, \frac{q_{\rho} \, p_{\sigma}}{2 \, p\cdot q}  , 
\quad  \tilde{P}_L^{\mu\nu}= \frac{4 \, x^2}{Q^2} p^{\mu} \, p^{\nu}  , \quad
\tilde{P}_5^{\mu\nu}= \frac{1}{2(1-\epsilon)} \,  \left[ -g^{\mu\nu} + \Tilde{P}_L^{\mu\nu} \right] .
\end{eqnarray}
The functions $g_i^{jk}$ contain collinear divergences, that appear as poles in $\epsilon$ (at NLO, simple poles in $\epsilon$).  
We factorize these divergences 
using the  $\overline{\mathrm{MS}}$ scheme, i.e., removing the quantities
\begin{align}
&\tilde{g}_i^{qq} (x,z) =  \Delta C_{o_i} \left[  -\frac{1}{\epsilon} + \gamma_E - \log(4 \pi) \right] [  \Delta f_{qq} (x) \delta(1-z) +   
			    P_{qq} (z)  \delta (1-x) ] , \nonumber \\
&\tilde{g}_i^{qg}(x,z)= \Delta C_{o_i} \,  \left[ -\frac{1}{\epsilon} + \gamma_E - \log(4\pi) \right]  \Delta P_{qg}(x)   \, \delta(1-z), \nonumber \\
 &\tilde{g}_i^{gq}(x,z)= \Delta C_{o_i} \,  \left[ -\frac{1}{\epsilon} + \gamma_E - \log(4\pi) \right]    P_{gq}(z)   \,  \delta(1-x) ,
\label{eq:gi_msbar}
\end{align} 
where $P_{jk}$ and $\Delta P_{jk}$ are the unpolarized and polarized LO Altarelli-Parisi splitting functions \cite{AltarelliParisi} and 
$\gamma_E=0,5772...$  is the Euler constant.
The quantity $\Delta f_{qq}$ is defined below in Eq. \eqref{eq:msbar_pol_fqq}. 
The finite functions obtained after factorization are the coefficients $\Delta C_i^{jk}$.

The main feature of the calculation described above is the correct use of $\gamma_5$ and the Levi-Civita tensor appearing in equations 
\eqref{eq:pol_fermion} and \eqref{eq:pol_gluon}, which  is not straightforward in $d\neq 4$ dimensions.
For our calculations we use the original prescription of `t Hooft and Veltman \cite{HV}, afterwards systematized by Breitenlohner and Maison \cite{BM} 
(HVBM scheme), which is the most reliable and consistent scheme \cite{BM,Consistency}.

In the HVBM scheme explicit definitions for $\gamma_5$ and 
$\epsilon_{\mu\nu\rho\sigma}$ are given. In particular, 
\begin{align}
 \gamma_5= \frac{i}{4!}\, \epsilon^{\mu\nu\rho\sigma} \, \gamma_{\mu}\gamma_{\nu}\gamma_{\rho}\gamma_{\sigma} \, \, ,
\label{eq:HBVM_gamma5_def}
\end{align}
and the $\epsilon$-tensor is regarded as a genuinely four-dimensional object with its components vanishing in all unphysical dimensions. 
The $d$-dimensional Minkowski space is then explicitly divided into two subspaces, a four-dimensional one and a $(d-4)$-dimensional one, 
each of them equipped with its metric tensor, $\tilde{g}$ and $\hat{g}$ respectively.
The Dirac matrices are also split into a four-dimensional and a $(d-4)$-dimensional part:
\begin{align}
\gamma_{\mu}=\tilde{\gamma}_{\mu}+\hat{\gamma}_{\mu} \, \, .
\label{eq:HBVM_gamma5}
\end{align}
Each part satisfies the usual anticommutation relation.

As defined above, the $\gamma_5$ anticommutes with $\tilde{\gamma}_{\mu}$ but commutes with  $\hat{\gamma}_{\mu}$. 
As a result, besides the usual scalar products $p' \cdot p'$ ($p'$ being the moment of the outgoing parton), 
products $\hat{p}'\cdot\hat{p}'$ show up in calculations, with $\hat{p}'$ the $(d-4)$-dimensional component of the momentum $p'$.
Every time this type of products appear, the two-particle phase space in Eq. \eqref{eq:partonicgi} must be used in the less integrated form \cite{VogelsangGluonic}
\begin{align}
d\Gamma &= \frac{1}{8 \pi} \;  \left( {4 \pi} \right)^{\epsilon} \; \frac{-\epsilon}{\Gamma(1-\epsilon)} \; 
	    \int_0^{sz(1-z)} d(\hat{p'}^2) \, ( \hat{p'}^2 )^{-(1+\epsilon)}.
\label{eq:ps22_HVBM}
\end{align}
For those terms which are independent of $\hat{p'}^2$ the integration is trivial and one recovers the result of \cite{AltarelliSIDIS}.

When computing the amplitudes, $\gamma_{5}$ matrices are present not only due to the helicity projectors, but also because of the weak interaction vertex. 
With the definition of $\gamma_5$ given in the HVBM scheme, it is important to consistently define the couplings with the quiral fields. 
It is shown in Ref. \cite{Vertix} that the correct $W^-$ vertex is obtained through the  symmetrization 
\begin{align}
 \gamma^{\mu} \, (1-\gamma_5) \, \to \, \frac{1}{2}(1+\gamma_5) \, \gamma^{\mu} \, (1-\gamma_5) = \tilde{\gamma}^{\mu} \, (1-\gamma_5) \, \, .
\label{eq:HBVM_vertice_W}
\end{align}
Analogously, the $Z$ vertex is obtained by
\begin{align}
 \gamma^{\mu}(g_V^q -g_A^q \, \gamma_5) &= \gamma^{\mu}\, (g_V^q - g_A^q) + g_A^q \,  \gamma^{\mu} \, (1 - \gamma_5) 
\to \,   \gamma^{\mu} \, (g_V^q - g_A^q) + g_A^q \, \tilde{\gamma}^{\mu} \, (1-\gamma_5) \, \, .
\label{eq:HBVM_vertice_Z}
\end{align}

Finally, an important property of the HVBM prescription for $\gamma_5$ is that it leads to helicity non-conservation at the $qqg$ vertex in $d$ dimensions, 
expressed by a non-vanishing difference of unpolarized and polarized $d$-dimensional LO quark-to-quark splitting functions,
\begin{align}
\Delta P_{qq}^{d=4-2\epsilon}(x)-P_{qq}^{d=4-2\epsilon}(x)= 4 \, C_F \, \epsilon \,  (1-x) \, \, . 
\label{eq:deltaPqq-Pqq}
\end{align}
As it is discussed in \cite{MertigVogelsang}, this result entails some disagreeable consequences, 
such as  non-con\-ser\-va\-tion of the flavour non-singlet axial current and 
an incorrect result for the ${\cal O}(\alpha_s)$ correction to the Bj\o rken sum rule. 
It is then customary to slightly modify the definition of the $\overline{\rm{MS}}$ scheme in the polarized case and define the quantity 
\begin{align}
\Delta f_{qq}(x) = \Delta P_{qq}(x) +4 C_F \, \epsilon (1-x)
\label{eq:msbar_pol_fqq}
\end{align}
to be removed in Eq. \eqref{eq:gi_msbar}.

We present now the NLO coefficients $\Delta C_i^{jk}$.  All the traces have been computed with the program {\sc Tracer} \cite{tracer}, 
which masters most of the intricacies of HVBM scheme. All results are given in the $\overline{\mathrm{MS}}$ scheme, as defined above. 
Here, $\mu_F^{IS}=\mu_F^{FS}=\mu_F$ is made, with $\mu_F^{IS}$ and $\mu_F^{FS}$ the initial state and final state factorization scales respectively.
For the sake of brevity we suppress in the following results the argument $(x,z)$ of the coefficient functions. For $i=1$, 
\begin{eqnarray}
\Delta C_1^{qq,V} &=&  C_1^{qq,V} - \lambda_V^V \, C_F \, 2 \, (1-x) \, (1-z) ,\\
\Delta C_1^{qg,V} &=& \lambda_V^V \,  \frac{1}{2} \, \left\{ \delta (1-z) \, \left[\Delta \tilde P_{qg}(x) \,  \log \left(\frac{Q^2}{\mu_F ^2} \, \frac{1-x}{x} \right) 
+2 \,  (1-x) \right] \right. \nonumber \\ 
&& \ \ \left. + \Delta \tilde P_{qg}(x) \, \left[ \frac{1}{(1-z)_+}+\frac{1}{z}-2 \right] \right\} , \\
\Delta C_1^{gq,V} &=& C_1^{gq,V} -  \lambda_V^V \, C_F \, 2\, (1-x) \, z ;
\end{eqnarray}
for $i=5$
\begin{eqnarray}
\Delta C_5^{qq,V} = \frac{\lambda_A^V  }{\lambda_V^V} \, C_1^{qq,V}, \quad
\Delta C_5^{qg,V} = \lambda_A^V \, \left[ \frac {\Delta C_1^{qg,V} } {\lambda_V^V} - \Delta P_{qg} (x)  2 \, \frac{1-z}{z}   \right] , \quad
\Delta C_5^{gq,V} = \frac{\lambda_A^V  }{\lambda_V^V} \, C_1^{gq,V};
\end{eqnarray}
and finally, the longitudinal coefficients are
\begin{eqnarray}
\Delta C_L^{qq,V} =  \frac{\lambda_A^V }{\lambda_V^V}  \, C_L^{qq,V}, \quad
\Delta C_{L}^{qg,V} = 0, \quad
\Delta C_L^{gq,V} =  \frac{\lambda_A^V }{\lambda_V^V}  \, C_L^{gq,V}. 
\end{eqnarray}
The functions $C_i^{jk}$ are the unpolarized coefficients and are shown in Appendix \ref{sec:ap_SIDIS_Nopol}. 
The factors $\lambda_A^V$ and $\lambda_V^V$ are given in Eq. \eqref{eq:lambdaV}. 
The quantity $\Delta \tilde{P}_{qg}$ is defined for simplicity as 
\begin{equation}
\Delta \tilde{P}_{qg}(x)=2 \, \Delta {P}_{qg}(x) = 2\,x -1.
\label{eq:TildeAP}
\end{equation}
Finally, the coefficients $\Delta C_4^{jk}$ can be obtained as
\begin{equation}
\Delta C_4^{jk,V} = \Delta C_L^{jk,V} + 2x\,\Delta C_5^{jk,V} .
\end{equation}

We note that the results for the electromagnetic case ($\Delta C_i^{jk,\gamma}$)  are in agreement with those of Ref. \cite{deFlorianLambda}.

\vspace{0.5cm}

\noindent Finally, we make some comments on the unpolarized case. 
It is convenient to define 
\begin{equation}
({\mathscr F}_1,{\mathscr F}_2,{\mathscr F}_3)=(2F_1,F_2/x,F_3)
\label{eq:F}
\end{equation}
and
\begin{equation}
{\mathscr F}_L= {\mathscr F}_2 - {\mathscr F}_1=F_L/x,
\end{equation}
and thus, the structure functions can be written at NLO as
\begin{align}
{\mathscr F}_i^{H,V}(x,z,Q^2)&=\sum_{q_a,q_b} \xi_i^{q_a} \left\{C_{o_i}^V  D_{q_b}^H(z,Q^2) \, q_a(x,Q^2) + \frac{\alpha_s(Q^2)}{2\pi} \left[ D_{q_b}^H  \otimes C_i^{qq,V} \otimes q_a \right. \right. \nonumber \\  
& \ \ \left. \left. + D_g^H \otimes C_i^{gq,V} \otimes q_a  + D_{q_b}^H \otimes  C_i^{qg,V} \otimes g  \right](x,z,Q^2) \right\},
\label{eq:sfun_nopol}
\end{align}
with 
\begin{equation}
  C_{o_1}^{V} = \lambda_V^V \quad ;  \quad  C_{o_3}^{V} = -\lambda_A^V \quad ; \quad C_{o_L}^{V} = 0. 
\end{equation}
The factor  $\xi_i^{q_a}$ takes the value $\xi_1^{q_a}=1$ for every $q_a$, 
while it is $\xi_3^{q_a}= \xi_L^{q_a}=1$ if $q_a$ refers to a quark and $\xi_3^{q_a}= \xi_L^{q_a}=-1$ for $q_a$ representing an anti-quark.

The computation of these coefficients is similar to the polarized ones. In this case, the spin-averaged amplitude must be used, and the projectors are
\begin{eqnarray}
P_L^{\mu\nu}= \frac{8\, x^2}{Q^2} \; p^{\mu} \, p^{\nu},\quad 
P_1^{\mu\nu}= \frac{1}{1-\epsilon} \left( -g^{\mu\nu} + \frac{1}{2} P_L^{\mu\nu} \right) , \quad
P_3^{\mu\nu}= i\; \epsilon^{\mu\nu\rho\sigma} \; \frac{q_{\rho} \, p_{\sigma}}{p \cdot q}.
\end{eqnarray}
The same care as in the polarized case must be taken when dealing with the $\gamma_5$ matrices present in the weak vertices. 
All NLO $\overline{\mathrm{MS}}$ coefficient functions $C_{i}^{jk,V}$ are collected in Appendix \ref{sec:ap_SIDIS_Nopol}.

\section{Structure functions at an EIC  \label{sec:phenomenology}}

\begin{figure}[h!tbp]
\begin{center}
\epsfig{file=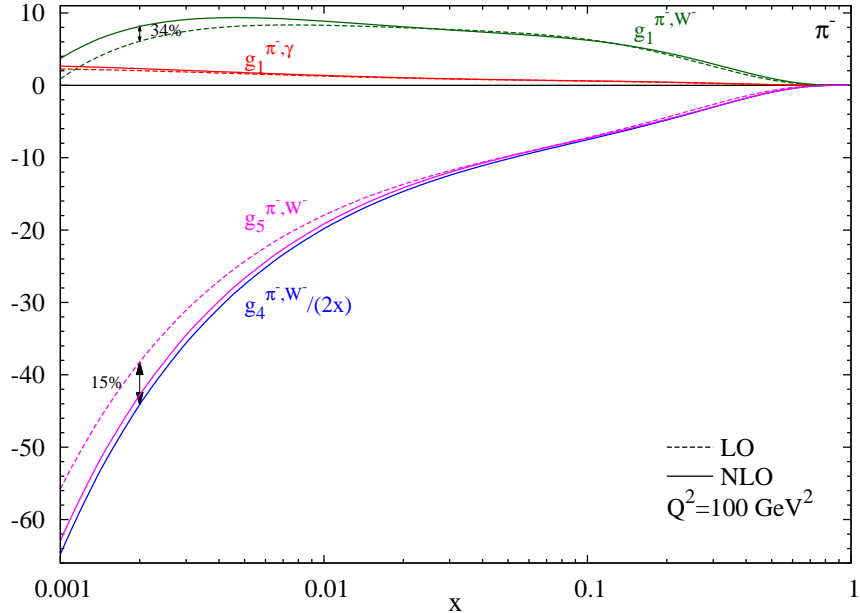,width=0.7\textwidth}
\end{center}
\caption{CC spin dependent SIDIS structure functions $g_1^{\pi^-,W^-}$, $g_5^{\pi^-,W^-}$, and $g_4^{\pi^-,W^-}/(2x)$, at $Q^2 = 100\, {\rm GeV}^2$ 
for the $z$-bin $z=0.2-0.4$. The dashed lines show the LO results (the one for $g_4^{\pi^-,W^-}/(2x)$
coincides with that for $g_5^{\pi^-,W^-}$), while the solid curves are NLO. For comparison, we also show the electromagnetic $g_1^{\pi^-,\gamma}$.
}
\label{fig:LO&NLO}
\end{figure}

In this section we analyse the relevance of the NLO corrections to the SIDIS structure functions we have computed. 
We focus on the CC case, i.e. the interaction via a $W^-$ boson, since it allow us to  achieve a full flavour separation.
These functions will be accessible in a future EIC, whose  characteristics can be found in Ref. \cite{EIC}. 
We will study the behaviour of the structure functions for an energy scale of $Q^2=100 \, {\rm GeV}^2$, for which the $x$-range foreseen is approximately 
$4.10^{-3}<x<1$ (see Fig. 7.16 of Ref. \cite{EIC}). 
We will consider four bins in the hadronic $z$ variable, with $0.1<z<0.8$ and rely on the DSS fragmentation functions set of \cite{DSS}.

In Fig. \ref{fig:LO&NLO} we show the spin dependent structure functions 
$g_1^{\pi^-,W^-}$, $g_5^{\pi^-,W^-}$, and $g_4^{\pi^-,W^-}/(2x)$, at \mbox{$Q^2 = 100 \, {\rm GeV}^2$} for the bin $z=0.2-0.4$, 
using the DSSV set of pPDFs \cite{DSSV2}. Results are shown both at LO (dashed) and NLO (solid).
It is important to mention that NLO pPDFs are used also at LO, in order to pick up only the effect of the corrections introduced by the NLO coefficients. 
One observes that the NLO results differ from the LO description, particularly in the small-$x$ range. 
At $x=0.002$, for instance, the discrepancies are of about $34\%$ for $g_1$ and $15\%$ for $g_4/(2x)$
(these being larger than those for $g_5$ for all $x$).
Thus, the NLO description is crucial for a precise extraction of the pPDFs in the small-$x$ region, 
which is particularly interesting since it could be measured at the EIC with unprecedented precision. 
For comparison, we also show the electromagnetic $g_1^{\pi^-,\gamma}$, for which the NLO corrections are more moderated (below $15\%$ in the mentioned $x$ value).

\begin{figure}[!ht]
\begin{center}
\epsfig{file=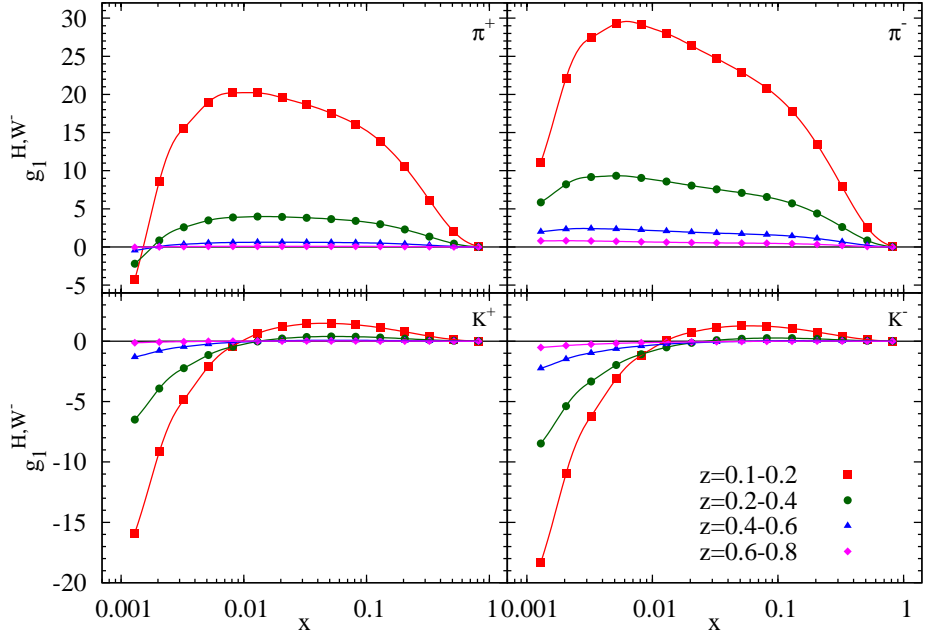,width=0.7\textwidth}
\end{center}
\caption{CC spin dependent structure function $g_1^{H,W^-}$ for $H=\pi^{\pm},K^{\pm}$ at $Q^2 = 100 \ {\rm GeV}^2$ for four different $z$-bins.}
\label{fig:g1w}
\end{figure}

\begin{figure}[!ht]
\begin{center}
\epsfig{file=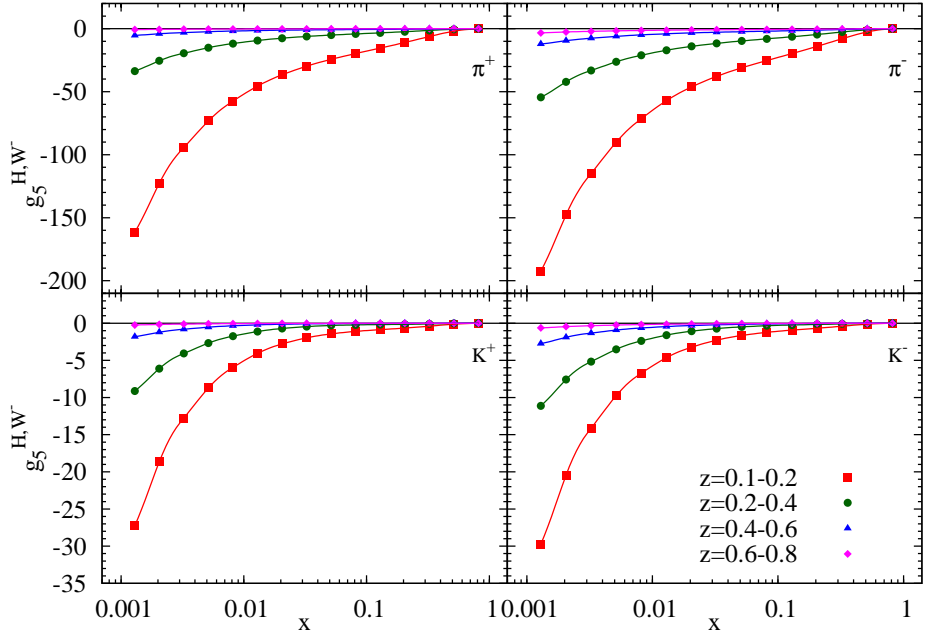,width=0.7\textwidth}
\end{center}
\caption{CC spin dependent structure function $g_5^{H,W^-}$ for $H=\pi^{\pm},K^{\pm}$ at $Q^2 = 100 \ {\rm GeV}^2$ for four different $z$-bins.}
\label{fig:g5w}
\end{figure}

In Fig. \ref{fig:g1w} we present the prediction for the NLO SIDIS structure functions $g_1^{H,W^-}$  
with $H=\pi^{\pm},K^{\pm}$ for the four $z$-bins considered. In all cases, the DSSV set of pPDFs where used.
We can understand the behaviour of these structure functions, at least qualitatively, by considering the behaviour of pPDFs and FFs, 
taking into account that only the four channels of Eq. \eqref{eq:channels} must be considered, as is discussed next.

We focus now on the $\pi^-$ case. 
The production of this hadron in the final state is mostly due to the hadronization of 
a quark $d$ (between $30\%$ and $46\%$) or a quark $\bar{u}$ (between $28\%$ and $41\%$). This means that the quark present in the initial state must be a quark $u$ or $\bar{d}$ respectively.
Thus, the behaviour of the structure function is dominated by that of the $\Delta u$ and $\Delta \bar{d}$ pPDFs, 
the first of them being positive for all $x$-range, and the last one negative. 
Given that the grater contribution is that of the $u$ channel, the $\pi^-$ structure function $g_1^{\pi^-,W^-}$ is always positive.
For $\pi^+$, however, at low $x$ contributions from $\Delta \bar{d}$ and $\Delta \bar{s}$ (negative both) are larger 
and so $g_1^{\pi^+,W^-}$ becomes negative.

 On the other hand, given the magnitude of the fragmentation functions in DSS set\footnote{The same being valid for other sets of FFs \cite{AKK1}.}, 
the $K^-$ production for low $z$ is dominated by the hadronization of a quark $\bar{c}$ (around $60\%$), 
meaning that the process is initiated by an $\bar{s}$ quark.
Thus, the shape of $g_1^{K^-,W^-}$ is practically that of the $\Delta\bar{s}$ distribution, with a sign change in the DSSV set around $x \sim 0.02 $.
For $0.4<z<0.6$ the $\bar{c}$ and $\bar{u}$ hadronization probabilities are comparable, and for $0.6<z<0.8$ the last one is even larger,
but cancellations between $\Delta u$ and $\Delta \bar{d}$ occur such that the behaviour of $g_1^{K^-,W^-}$ remains very similar to that just described.

In Fig. \ref{fig:g5w} we show the prediction for the SIDIS structure functions $g_5^{H,W^-}$  at NLO. 
By looking at Eq. \eqref{eq:sfun_LO_CC} we can note that the main difference for these structure functions 
with respect to the previous discussion relies in the fact that these ones depend on $(-\Delta u)$, $\Delta \bar{d}$ and $\Delta \bar{s}$, all of them basically negative
for all $x$-range (except for $\Delta \bar{s}$ that is positive for large $x$, but that effect is not noticeable).
That explains why $g_5^{H,W^-}$ is negative for all hadrons considered ($H=\pi^\pm,K^\pm$) and for all $x$ values.

\begin{figure}[!t]
\begin{center}
\epsfig{file=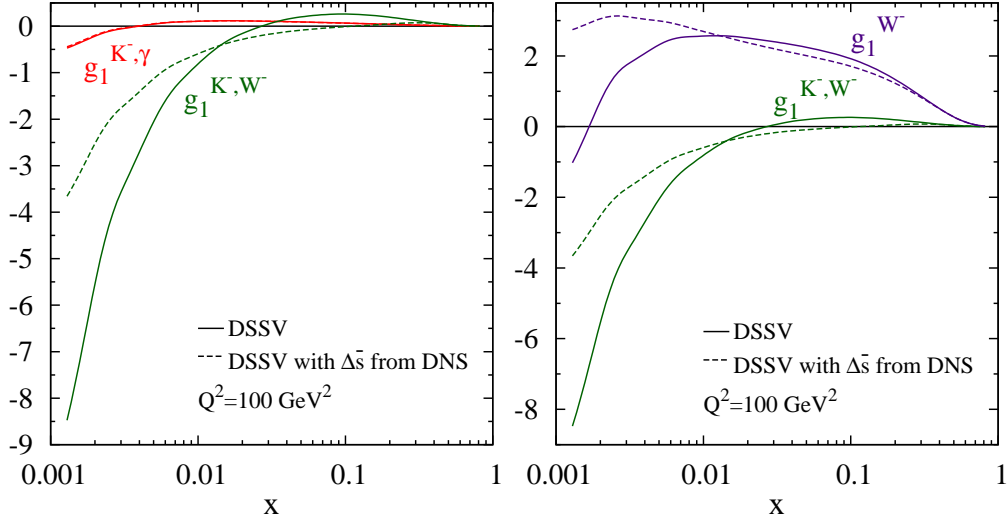,width=0.8\textwidth}
\end{center}
\caption{Left: CC and electromagnetic spin dependent SIDIS structure functions $g_1^{K^-,W^-}$ (green) and $g_1^{K^-,\gamma}$ (red) for $z=0.2-0.4$. 
We present the results obtained using DSSV pPDFs set (solid) and all pPDFs of DSSV set except for $\Delta \bar{s}$, taken from DNS set (dashed). 
Right: The same, for CC spin dependent SIDIS structure functions $g_1^{K^-,W^-}$ for $z=0.2-0.4$ (green) 
and DIS structure function $g_1^{W^-}$ (violet).
}
\label{fig:DNSs_K-}
\end{figure}

As it was explained above, the $g_1^{K-,W-}$ structure function behaviour is closely related to the $\Delta \bar{s}$ distribution. 
Thus, one can expect it to be sensitive to a small change in that pPDF. 
We show in the left hand side of  Fig. \ref{fig:DNSs_K-} the CC and electromagnetic structure functions $g_1^{K^-,W^-}$ (green) 
and $g_1^{K^-,\gamma}$ (red) calculated with the DSSV set of pPDFs (solid)
and the same result after modifying {\it only} the $\Delta \bar{s}$ distribution according to the one in the DNS set \cite{DNS} (dashed), as a way to emphasize the sensitivity of the observable on that flavour.
We note large discrepancies between both sets of pPDFs in the CC case, and even observe a region in which they have opposite signs, unlike the electromagnetic case, 
for which the difference between both sets is much smaller. 
In the last case, the presence of an $\bar{s}$ quark in the initial state implies (at LO) the hadronization of an $\bar{s}$ quark, 
but its FF into a $K^-$ hadron is almost negligible.
The sensitivity of $g_1^{K-,W-}$ is also stronger than that of the DIS structure function $g_1^{W^-}$ for some $x$-ranges 
as it can be observed in the right hand side of Fig. \ref{fig:DNSs_K-}.

\begin{figure}[!t]
\begin{center}
\epsfig{file=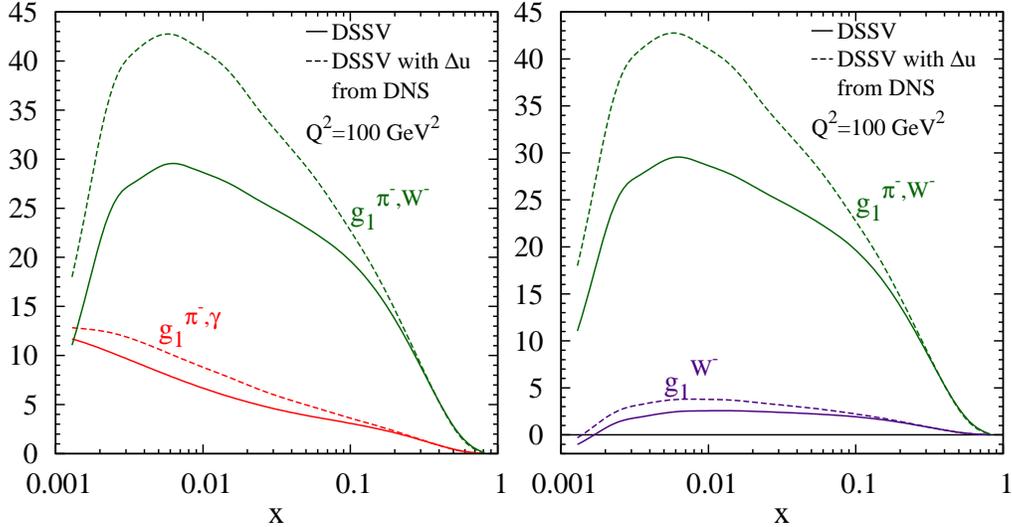,width=0.8\textwidth}
\end{center}
\caption{Left: CC and electromagnetic spin dependent structure functions $g_1^{\pi^-,W^-}$ (green) and $g_1^{\pi^-,\gamma}$ (red) respectively, 
calculated using DSSV pPDFs set (solid) and DSSV set except for $\Delta u$, which is taken from DNS set (dashed). In both cases, we use $z=0.1-0.2$. 
Right: The same for CC spin dependent SIDIS structure functions $g_1^{\pi^-,W^-}$ for $z=0.1-0.2$ (red) and DIS structure function $g_1^{W^-}$ (violet).
}
\label{fig:DNSu_Pi-}
\end{figure}

Similarly, one can expect the $g_1^{\pi^-,W^-}$ structure function to be particularly sensitive to a variation of $\Delta u$. 
We show in the left hand side of Fig. \ref{fig:DNSu_Pi-} the CC and electromagnetic spin dependent structure functions $g_1^{\pi^-,W^-}$ (green) 
and $g_1^{\pi^-,\gamma}$ (red) respectively, 
calculated using DSSV pPDFs set (solid) and DSSV set with the exception of $\Delta u$, taken from DNS set (dashed). 
By comparing these two plots, we can note that in fact the CC SIDIS structure function is much more sensitive to $\Delta u$ than the electromagnetic one, 
and thus better flavour separation can be achieved from such measurement. 

  We also compare in the right hand side of Fig. \ref{fig:DNSu_Pi-} the CC spin dependent SIDIS structure functions 
$g_1^{\pi^-,W^-}$ for $z=0.2-0.4$ (red) and DIS structure function $g_1^{W^-}$ (violet) obtained using DSSV pPDFs set (solid) 
and DSSV set with the exception of $\Delta u$, taken from DNS set (dashed). 
The sensitivity of these functions on $\Delta u$ are similar, but one can still profit from the larger structure function in the SIDIS case.

\section{Conclusions \label{sec:conclusions}}

We have computed the next-to-leading order QCD corrections to the complete set of 
polarized semi-inclusive electroweak structure functions in a consistent scheme. 

We have performed a phenomenological study of possible measurements in an EIC.
The NLO corrections are found to be important in the small-$x$ range, 
the kinematical region of particular interest at a high energy collider.

We analysed the sensitivity of SIDIS CC structure functions on different pPDFs, 
and showed that they provide an excellent tool to disentangle the full set of flavour spin dependent distributions with an unprecedent 
precision by including the electroweak polarized SIDIS structure functions.
In particular, we explicitly show that the structure function $g_1^{K-,W-}$ is sensitive to the very poorly known $\Delta \bar{s}$ distribution  
while the function $g_1^{\pi-,W-}$ is particularly sensitive to $\Delta u$.

\noindent {\bf Acknowledgements.}

This work was supported in part by UBACYT, CONICET, ANPCyT and the Research Executive Agency (REA) of the European Union under the Grant Agreement number PITN-GA-2010-264564 (LHCPhenoNet).

\renewcommand\appendixname{\rm{Appendix} \arabic{appendix}}

\appendix

\section{Unpolarized SIDIS Coefficient Functions \label{sec:ap_SIDIS_Nopol}}

%

%
Here we list all unpolarized coefficients $C_{i}^{jk,V}$ for SIDIS as introduced in 
Section \ref{chap:NLO}. To keep the expressions as short as possible we define
\begin{eqnarray}
&&\tilde{P}_{qq} (\xi)= \frac{1}{{\rm C_F}} P_{qq} \; , \quad
\tilde{P}_{qg}(\xi)=2 \, {P}_{qg}(\xi) \; , \quad
\tilde{P}_{gq}(\xi)=  \frac{1}{{\rm C_F}} {P}_{gq}(\xi) \;, \nonumber \\
&& \quad  L_1(\xi)=(1+\xi^2)\left(\frac{\log (1-\xi)}{1-\xi}\right)_+ \; , \quad
L_2(\xi)=\frac{1+\xi^2}{1-\xi}\log \xi\;\;.
\label{eq:Definitions}
\end{eqnarray}
In what follows, we always suppress the argument $(x,z)$ of the 
coefficient functions.
All results presented here are given in the $\overline{\mathrm{MS}}$
scheme. 

The unpolarized coefficients $C_{i}^{jk,V}$ for SIDIS are
\begin{eqnarray}
\label{eq:c_nopol_first}
C_{1}^{qq,V} &=& \lambda_V^V \; C_F \Bigg[ -8 \, \delta(1-x)\delta(1-z) \nonumber \\
&&
+\delta(1-x) \left[ \tilde{P}_{qq}(z) \log \left( \frac{Q^2}{\mu_F^2} \right) +
 L_1(z)+L_2(z)+(1-z)\right]  \nonumber  \\
&&
+\delta(1-z) \left[ \tilde{P}_{qq}(x) \log\left( \frac{Q^2}{\mu_F^2} \right) +
L_1(x)-L_2(x)+(1-x)\right] \nonumber \\ 
&&+ 2\frac{1}{(1-x)_+}\frac{1}{(1-z)_+}- \frac{1+z}{(1-x)_+}-
\frac{1+x}{(1-z)_+}+2(1+x\,z) \Bigg] , \\  
C_1^{qg,V} &=& \lambda_V^V \; \frac{1}{2} \Bigg[ \delta (1-z) \left[\tilde{P}_{qg}(x)
\log \left( \frac{Q^2}{\mu_F^2} \frac{1-x}{x} \right) +2x(1-x)\right] \nonumber \\
&& + \tilde{P}_{qg}(x) \left\{ \frac{1}{(1-z)_+}+\frac{1}{z}-2\right\} \Bigg]  , \\
C_1^{gq,V} &=& \lambda_V^V \; C_F \Bigg[ \tilde{P}_{gq}(z) \left(\delta (1-x) \log\left(
\frac{Q^2}{\mu_F^2}z(1-z)\right)+\frac{1}{(1-x)_+}\right) \nonumber \\ 
&&+  z \, \delta(1-x)+2(1+x-x\,z)-\frac{1+x}{z}\Bigg] ; 
\end{eqnarray}
\begin{eqnarray}
C_L^{qq,V} &=&  \lambda_V^V \, 4 \; C_F x z ,\\
C_L^{qg,V} &=&  \lambda_V^V \, 4 \; x(1-x) , \\
C_L^{gq,V} &=&  \lambda_V^V \, 4 \; C_F x (1-z) ;
\end{eqnarray}
\begin{eqnarray}
C_3^{qq,V} &=& -\lambda_A^V \; \left[ \frac{C_1^{qq,V}}{\lambda_V^V} -  C_F \, 2 \, (1-x)\,(1-z) \right] ,\\
C_3^{qg,V} &=& -\lambda_A^V \; \left[ \frac{C_1^{qg,V}}{\lambda_V^V} - 2 \, P_{qg}(x) \frac{1-z}{z} \right] , \\
C_3^{gq,V} &=& -\lambda_A^V \; \left[ \frac{C_1^{gq,V}}{\lambda_V^V} -  C_F \, 2 \, (1-x)\, z \right] \, \, .
\label{eq:c_nopol_last}
\end{eqnarray}

We note that all our results in Eqs. \eqref{eq:c_nopol_first} - \eqref{eq:c_nopol_last} are in agreement with Refs. \cite{AltarelliSIDIS,deFlorianLambda} .

\section{Unpolarized and polarized DIS Coefficient Functions \label{sec:ap_DIS}}

For the sake of completeness, we present the NLO unpolarized and polarized DIS strucutre functions, $F_i^{V}$ and $g_i^{V}$ respectively. 
These, can be expressed as convolutions of non-perturbative (p)PDFs with short-distance coefficients 
($C_{1,3,L}^{j,V}(x,Q^2)$ in the unpolarized case and $\Delta C_{1,5,L}^{j,V}(x,Q^2)$ in the polarized one).
At NLO, these read
\begin{equation}
{\mathscr F}_i^{V}(x,Q^2)= \frac{1}{2} \sum_{q_a}  \xi_i^{q_a} \left\{ C_{o_i}^V \, q_a(x,Q^2) 
		    + \frac{\alpha_s(Q^2)}{2\pi} \left[ C_i^{q,V} \otimes q_a +  C_i^{g,V} \otimes g  \right](x,Q^2)\right\} \,\, ,
\label{eq:sfun_nopol_DIS}
\end{equation}
\begin{equation}
g_i^{V}(x,Q^2)= \frac{1}{2} \sum_{q_a} \Delta \xi_i^{q_a} \left\{\Delta C_{o_i}^V   \Delta q_a(x,Q^2) 
		    + \frac{\alpha_s(Q^2)}{2\pi} \left[  \Delta C_i^{q,V} \otimes \Delta q_a  + \Delta C_i^{g,V} \otimes \Delta g  \right](x,Q^2)\right\}.
\label{eq:sfun_pol_DIS}
\end{equation}
All the relevant factors have been introduced in section \ref{sec:NLO}.

The unpolarized coefficients are
\begin{eqnarray}
C_L^{q,V}&=& \lambda_V^V \,  C_F  \,  2\, x , \nonumber \\
C_1^{q,V}&=& \lambda_V^V \, C_F \,  \left\{  \log\left( \frac{Q^2}{\mu_F^2}\right) \tilde{P}_{qq} (x) +
	     \delta(1-x)\left(-\frac{9}{2}-\frac{\pi^2}{3}\right)  \right. \nonumber \\  
	  & & \left. + L_1(x)-L_2(x)  -\frac{3}{2} \frac{1}{(1-x)_+} +3\right\} , \nonumber \\
C_3^{q,V}&=& - \lambda_A^V \, \left[ \frac{ C_1^{q,V} }{\lambda_V^V} - C_F (1-x)\right], \nonumber \\
C_L^{g,V}&=& \lambda_V^V \,   2 \,  x  \, (1-x)  , \nonumber \\
C_1^{g,V}&=& \lambda_V^V \, \frac{1}{2} \left\{ \tilde{P}_{qg}(x) \left[ \log \left(\frac{Q^2}{\mu_F^2} \frac{1-x}{x}\right) -1\right] 
	      + 2  x (1-x) \right\} , \nonumber \\
C_3^{g,V}&=& 0;
\end{eqnarray}
and the polarized ones,
\begin{eqnarray}
\Delta C_L^{q,V}&=& \frac{\lambda_A^V}{\lambda_V^V}  C_L^{q,V}  \nonumber \\
\Delta C_1^{q,V}&=& C_1^{q,V}   - \lambda_V^V  \, C_F \, (1-x) , \nonumber \\
\Delta C_5^{q,V}&=&  \frac{ \lambda_A^V    }{\lambda_V^V} \,C_1^{q,V} , \nonumber \\
\Delta C_L^{g,V}&=& 0, \nonumber \\
\Delta C_1^{g,V}&=& \lambda_V^V \, \frac{1}{2} \, \left\{  \Delta \tilde{P}_{qg} \left[ \log \left( \frac{Q^2}{\mu_F^2} \frac{1-x}{x} \right) -1 \right] 
		    +2 (1-x)  \right\}, \nonumber \\
\Delta C_3^{g,V}&=& 0 \,\, ,
\end{eqnarray}
with $L_1(x)$, $L_2(x)$ and $\tilde{P}_{jk}$ defined in Eq. \eqref{eq:Definitions} and $\Delta \tilde{P}_{jk}$ defined in Eq. \eqref{eq:TildeAP}.

All our results are in agreement with   Refs. \cite{AltarelliSIDIS,VogelsangGluonic,AltarelliDIS,deFlorianSpin}.

\bibliographystyle{unsrt}

\begin{thebibliography}{99}

\bibitem{review}
  M.~Burkardt, C.~A.~Miller and W.~D.~Nowak,
  Rept.\ Prog.\ Phys.\  {\bf 73} (2010) 016201; 
  D.~de Florian, R.~Sassot, M.~Stratmann and W.~Vogelsang,
  Prog.\ Part.\ Nucl.\ Phys.\  {\bf 67} (2012) 251.
  


\bibitem{DSSV1}
  D.~de Florian, R.~Sassot, M.~Stratmann and W.~Vogelsang,
  Phys.\ Rev.\ D {\bf 80} (2009) 034030.

\bibitem{DSSV2}
  D.~de Florian, R.~Sassot, M.~Stratmann and W.~Vogelsang,
  Phys.\ Rev.\ Lett.\  {\bf 101} (2008) 072001.


\bibitem{EIC} 
  D.~Boer, M.~Diehl, R.~Milner, R.~Venugopalan, W.~Vogelsang, D.~Kaplan, H.~Montgomery and S.~Vigdor {\it et al.},
  arXiv:1108.1713 [nucl-th].

\bibitem{EWSF}
  B.~Lampe,
  Phys.\ Lett.\ B {\bf 227} (1989) 469;
  W.~Vogelsang and A.~Weber,
  Nucl.\ Phys.\ B {\bf 362} (1991) 3;
  P.~Mathews and V.~Ravindran,
  Phys.\ Lett.\ B {\bf 278} (1992) 175.

\bibitem{AltarelliSIDIS} 
  G.~Altarelli, R.~K.~Ellis, G.~Martinelli and S.~-Y.~Pi,
  Nucl.\ Phys.\ B {\bf 160}, 301 (1979).

\bibitem{deFlorianLambda} 
  D.~de Florian, M.~Stratmann and W.~Vogelsang,
  Phys.\ Rev.\ D {\bf 57}, 5811 (1998).


\bibitem{Anselmino}
  M.~Anselmino, P.~Gambino and J.~Kalinowski,
  Z.\ Phys.\ C {\bf 64} (1994) 267.

\bibitem{PDG}
  J.~Beringer {\it et al.}  [Particle Data Group Collaboration],
  Phys.\ Rev.\ D {\bf 86} (2012) 010001.

\bibitem{fracture}
  L.~Trentadue and G.~Veneziano, 
  Phys.\ Lett.\ B {\bf 323}, 201 (1994);
   D.~Graudenz,
  Nucl.\ Phys.\ B {\bf 432} (1994) 351;
  D.~de Florian, C.~A.~Garcia Canal and R.~Sassot,
  Nucl.\ Phys.\ B {\bf 470} (1996) 195.



\bibitem{Callan} 
  C.~G.~Callan, Jr. and D.~J.~Gross,
  Phys.\ Rev.\ Lett.\  {\bf 22}, 156 (1969).

\bibitem{regularization}
  C.~G.~Bollini and J.~J.~Giambiagi,
  Nuovo Cim.\ B {\bf 12} (1972) 20.


\bibitem{HV}
  G.~'t Hooft and M.~J.~G.~Veltman,
  Nucl.\ Phys.\ B {\bf 44} (1972) 189.


\bibitem{AltarelliParisi}
  G.~Altarelli and G.~Parisi,
  Nucl.\ Phys.\ B {\bf 126} (1977) 298.


\bibitem{BM}
  P.~Breitenlohner and D.~Maison,
  Commun.\ Math.\ Phys.\  {\bf 52} (1977) 11.


\bibitem{Consistency} 
  G.~Bonneau,
  Phys.\ Lett.\ B {\bf 96}, 147 (1980).


\bibitem{VogelsangGluonic} 
  W.~Vogelsang,
  Z.\ Phys.\ C {\bf 50}, 275 (1991).


\bibitem{Vertix}
  J.~G.~Korner, N.~Nasrallah and K.~Schilcher,
  Phys.\ Rev.\ D {\bf 41} (1990) 888.



\bibitem{MertigVogelsang}
  R.~Mertig and W.~L.~van Neerven,
  Z.\ Phys.\ C {\bf 70} (1996) 637;
  W.~Vogelsang,
  Phys.\ Rev.\ D {\bf 54} (1996) 2023;
  Nucl.\ Phys.\ B {\bf 475} (1996) 47.




\bibitem{tracer}
  M.~Jamin and M.~E.~Lautenbacher,
  Comput.\ Phys.\ Commun.\  {\bf 74} (1993) 265.




\bibitem{DSS} 
  D.~de Florian, R.~Sassot and M.~Stratmann,
  Phys.\ Rev.\ D {\bf 75}, 114010 (2007).




\bibitem{AKK1}
  S.~Albino, B.~A.~Kniehl and G.~Kramer,
  Nucl.\ Phys.\ B {\bf 725} (2005) 181.


\bibitem{DNS}
  D.~de Florian, G.~A.~Navarro and R.~Sassot,
  Phys.\ Rev.\ D {\bf 71} (2005) 094018.


\bibitem{AltarelliDIS}
  G.~Altarelli, R.~K.~Ellis and G.~Martinelli,
  Nucl.\ Phys.\ B {\bf 157} (1979) 461.



\bibitem{deFlorianSpin}
  D.~de Florian and R.~Sassot,
  Phys.\ Rev.\ D {\bf 51} (1995) 6052.




\end{thebibliography}

\end{document}